\documentclass[fleqn,twoside,psfig]{article}
\usepackage{espcrc2}

\usepackage{psfig}
\usepackage[figuresright]{rotating}

\newcommand{\AmS}{{\protect\the\textfont2
  A\kern-.1667em\lower.5ex\hbox{M}\kern-.125emS}}

\newcommand{\eff}{\varepsilon}
\newcommand{\ra}{\rightarrow}
\newcommand{\piz}{\pi^0}

\newcommand{\hq}{h4\pi^0}

\newcommand{\thd}{3h2\pi^0}
\newcommand{\tht}{3h3\pi^0}

\newcommand{\clsi}{$e$}
\newcommand{\clsii}{$\mu$}
\newcommand{\clsiii}{$h$}
\newcommand{\clsiv}{$h\pi^0$}
\newcommand{\clsv}{$h2\pi^0$}
\newcommand{\clsvi}{$h3\pi^0$}
\newcommand{\clsvii}{$3h$}
\newcommand{\clsviii}{$3h\pi^0$}
\newcommand{\clsix}{$3h2\pi^0$}
\newcommand{\clsx}{$3h3\pi^0$}
\newcommand{\clsxi}{$5h$}
\newcommand{\clsxii}{$5h\pi^0$}
\newcommand{\clsxiii}{$h4\pi^0$}

\newcommand{\phyi}{$e$}
\newcommand{\phyii}{$\mu$}
\newcommand{\phyiii}{$\pi^-$}
\newcommand{\phyiv}{$\pi^-\pi^0$}
\newcommand{\phyv}{$\pi^-2\pi^0$}
\newcommand{\phyvi}{$\pi^-3\pi^0$}
\newcommand{\phyvii}{$\pi^-\pi^-\pi^+$}
\newcommand{\phyviii}{$\pi^-\pi^-\pi^+\pi^0$}
\newcommand{\phyix}{$\pi^-\pi^-\pi^+2\pi^0$}
\newcommand{\phyx}{$\pi^-\pi^-\pi^+3\pi^0$}
\newcommand{\phyxi}{$3\pi^-2\pi^+$}
\newcommand{\phyxii}{$3\pi^-2\pi^+\pi^0$}
\newcommand{\phyxiii}{$\pi^-4\pi^0$}

\newcommand{\bfg}{\begin{figure}}
\newcommand{\efg}{\end{figure}}
\newcommand{\bitm}{\begin{itemize}}
\newcommand{\eitm}{\end{itemize}}
\newcommand{\bnum}{\begin{enumerate}}
\newcommand{\enum}{\end{enumerate}}
\newcommand{\btbl}{\begin{table}}
\newcommand{\etbl}{\end{table}}
\newcommand{\btbu}{\begin{tabular}}
\newcommand{\etbu}{\end{tabular}}
\newcommand{\beq}{\begin{equation}}
\newcommand{\eeq}{\end{equation}}
\newcommand{\beqn}{\begin{eqnarray}}
\newcommand{\eeqn}{\end{eqnarray}}
\newcommand{\beqns}{\begin{eqnarray*}}
\newcommand{\eeqns}{\end{eqnarray*}}

\def\EPJC{{\it Eur. Phys. J.}}
\def\PL{{\it Phys. Lett.}}
\def\PR{{\it Phys. Rev.}}

\def\PRL{{\it Phys. Rev. Lett.}}

\def\CPC{{\it Comp. Phys. Comm.}}
\def\ea{{\it et al.}}
\def\Rt{$R_\tau$}

\def\br{branching ratio}

\title{Measurement of Branching Fractions in $\tau$ Decays}

\author{Michel Davier\address[LAL]{Laboratoire de l'Acc\'el\'erateur 
				   Lin\'eaire, 
				   IN2P3/CNRS-Universit\'e de Paris-Sud, \\ 
        		           BP34, 91898 Orsay, France}
 		and Changzheng Yuan\addressmark[LAL]\thanks{
	now at Institute of High Energy Physics, Beijing 100039, China}
	 \\
	{\it for the ALEPH Collaboration}}
       
\begin{document}

\begin{abstract}

  Full LEP-I data collected by the ALEPH detector during 1991-1995
running are analyzed in order to measure the $\tau$ decay 
branching fractions. The analysis follows the global method
used in the published study based on 1991-1993 data, with several 
improvements, especially concerning the treatment of 
photons and $\pi^0$'s. Extensive systematic studies are performed,
in order to match the large statistics of the data sample 
corresponding to 327148 measured and identified $\tau$ decays. 
Preliminary values for the branching 
fractions are obtained for the 2 leptonic channels and 11 hadronic
channels defined by their respective numbers of charged particles and 
$\pi^0$'s. Using previously published ALEPH results on final states
with charged and neutral kaons, corrections are applied so that 
branching ratios for exclusive final states 
without kaons are derived. Some physics implications of the results
are given, in particular concerning universality in the leptonic charged
weak current, isospin invariance in $a_1$ decays, and the separation of
vector and axial-vector components of the total hadronic rate.
\vspace{1pc}
\end{abstract}

\maketitle

\section{Introduction}

A complete and final analysis of $\tau$ decays is presented
using a global method. All data recorded at LEP-I with the ALEPH detector 
are used, thus providing an update of those previous results which were based 
on partial data sets. The increase in statistics ---the full sample
corresponds to about 2.5 times the luminosity used in the last published 
global analysis~\cite{aleph13_l,aleph13_h}--- not only allows for a reduction
of the dominant statistical error but, more importantly, provides a way to
better study possible systematic biases and to eventually correct for them.
Several improvements of the method have been introduced in order to achieve
a better control over the most relevant systematic uncertainties: 
simulation-independent measurement of the $\tau \tau$ 
selection efficiency, improved photon identification
especially at low energy where the separation between photons from $\tau$
decays and fake photons from fluctuations in hadronic or electromagnetic 
showers is delicate, a new method to correct the Monte Carlo simulation
for the rate of fake photons, and stricter criteria for channels with 
low branching fractions. For consistency and in order to maximally profit 
from the improved analysis all data sets recorded from 1991 to 1995 have
been reprocessed. The results presented in this paper thus supersede those
already published in Ref.~\cite{aleph01,aleph13_l,aleph13_h}.  
Only the measurements on final states containing kaons, which were 
already based on the full statistics, remain
unchanged~\cite{alephk3,alephks,alephkl,alephksum}.

\section{Experimental method}

\subsection{The data and simulated samples}

A detailed description of the ALEPH detector can be found
elsewhere~\cite{alephdet,alephperf}. 

Tau-pair events are simulated by means of a  Monte Carlo program
which includes initial state radiation computed
up to order $\alpha^2$ and exponentiated,
and  final state radiative corrections to order $\alpha$~\cite{was}.
The simulation of the subsequent $\tau$ decays also includes single photon
radiation for the decays with up to three hadrons in the final state.
The longitudinal spin correlation is taken into account~\cite{tauola}.
This simulation, with the detector acceptance
and resolution effects, is used to initially evaluate the corresponding
relative efficiencies and backgrounds. It also includes the tracking,
the secondary interactions of hadrons, bremsstrahlung and conversions.
For all these effects, detailed
comparisons with relevant data distributions are performed and corrections
to the MC-determined efficiencies are derived.

The data used in this analysis have been recorded at LEP I in 1991-1995.  
The numbers of detected $\tau$ decays are correspondingly 
132316 in 1991-1993 and 194832 in 1994-1995, for a total of about 
$3.3 \cdot 10^5$. The ratios between Monte Carlo and data statistics 
are 7.3 and 9.7 for 1991-1993 and 1994-1995 periods, respectively. 
Monte Carlo samples are generated for each year of data-taking in order
to follow as closely as possible the status of the detector components.

\subsection{Selection of $\tau \tau$ events}
\label{select}

The principal characteristics of $\tau \tau$ events in $e^+e^-$
annihilation are low multiplicity, back-to-back topology and missing
energy. Each event is divided into two hemispheres by an energy flow
algorithm~\cite{alephperf} which calculates all the visible energy avoiding
double-counting between the TPC and the calorimeter information. The jet
in a given hemisphere is defined by summing all the four-momenta of
all energy flow objects (charged and neutral). The energies in the two
hemispheres including the energies of photons from final state radiation,
$E_1$ and $E_2$, are useful variables for separating Bhabha, $\mu \mu$
and $\gamma \gamma$-induced events from the $\tau \tau$ sample, while
the relatively larger jet masses, wider opening angles, and higher 
multiplicities indicate $Z \rightarrow q \overline{q}$ events. 

All these features are incorporated in a standard selector used extensively
in ALEPH~\cite{aleph94,aleph13_l,aleph13_h}. 
In the previous analyses~\cite{aleph13_l,aleph13_h} additional cuts had
been introduced in order to further reduce the contamination from $ee$ and
$\mu\mu$ processes. In the present work it was chosen to simplify the 
procedure in order to conveniently measure selection efficiencies on the 
data, at the expense of a slightly larger background contamination which 
is anyway also measured in the data sample as explained later. 

We use the 'break-mix' method introduced for the
determination of the $\tau\tau$ cross section~\cite{aleph94} to measure the
efficiency of all the selection cuts. For every cut,
one hemisphere of the event is chosen judiciously so that it is unbiased
with respect to the cut under study and free of non-$\tau$ backgrounds. 
This procedure selects the opposite hemisphere as an unbiased 
$\tau$ decay which is then stored away. 
Pairs of selected hemispheres are combined to construct a $\tau\tau$ event
sample built completely from data. This sample is used to measure the 
efficiency of the given cut. 

The measured efficiencies are found to be very close to those 
obtained by the simulation, deviations being at most 
at the few per mille level. This situation stems from the facts that the
$\tau$ decay dynamics is ---apart from small branching ratio channels---
very well known, the selection efficiencies are large and the simulation
of the detector is adequate. The overall selection efficiency of $\tau\tau$
events is 78.8 \%. This value increases to 91.9 \% when
the $\tau\tau$ angular distribution is restricted to the detector polar
acceptance, giving a better indication for the efficiency of
the cuts designed to exclude non-$\tau\tau$ backgrounds. In addition, 
when expressed relatively to each $\tau$ decay, the
selection efficiencies are weakly dependent on the final state,
with a total relative span of only 10 \% for the 13 considered decay 
topologies.

\subsection{Estimation of non-$\tau\tau$ backgrounds}
\label{nontau}

A new method ---already implemented for the measurement of 
the $\tau$ polarization~\cite{alephpol}--- has been
developed to directly measure in the final data samples 
the contributions from the major non-$\tau$ backgrounds: Bhabhas, 
$\mu^+ \mu^-$ pairs, and $\gamma \gamma \rightarrow e^+ e^-, \mu^+ \mu^-$,
and hadrons events. The procedure does not require an absolute 
normalization from the Monte Carlo simulation of these channels,
only a qualitative description of the distribution of the discriminating
variables. The basic idea is to apply cuts on the data in order to reduce
as much as possible the $\tau \tau$ population while keeping a high
efficiency for the background source under study, {\it i.e.} the
reverse of what is done in the $\tau\tau$ selection.

The non-$\tau$ backgrounds in each channel are listed in Table~\ref{nobs_dt}. 
They amount to a total fraction of ($1.23 \pm 0.04$) \% 
in the full data sample.

\subsection{Charged particle identification}
\label{pid}

A 'good' track is defined to have a momentum greater than or equal 
to 0.10~GeV/c ,
$|\cos(\theta)| \le 0.95$, at least 4 hits in the TPC, and its
minimum distance to the interaction point within
2~cm transversally and 10~cm along the beams.
In classifying $\tau$ decays, only good tracks are used, after
removing those identified as electrons which are used to reconstruct
converted photons; the electrons identified as bad tracks
are also included in the reconstruction of conversions.

Charged particle identification is achieved with
a likelihood method incorporating the information from the relevant 
detectors. In this way, each charged particle is assigned a set of 
probabilities from which a particle type is chosen. No attempt is made 
in this analysis to separate kaons from pions in the hadron sample 
since final states containing
kaons have been previously studied~\cite{alephk3,alephks,alephkl}.

Eight discriminating variables are used in the identification procedure:
$dE/dx$ in the TPC, two estimators (transverse and longitudinal) of the 
shower profile in ECAL, the average shower width measured with the HCAL
tubes in the fired planes, the number of fired planes among the last ten,
the energy measured with HCAL pads, the number of hits in the muon
chambers ($\pm 4\sigma$-wide around the track extrapolation, where $\sigma$
is the standard deviation expected from multiple scattering), and finally,
the average distance (in units of the multiple-scattering standard deviation)
of the hits from their expected position in the muon chambers.

\begin{figure}[t]
   \centerline{\psfig{file=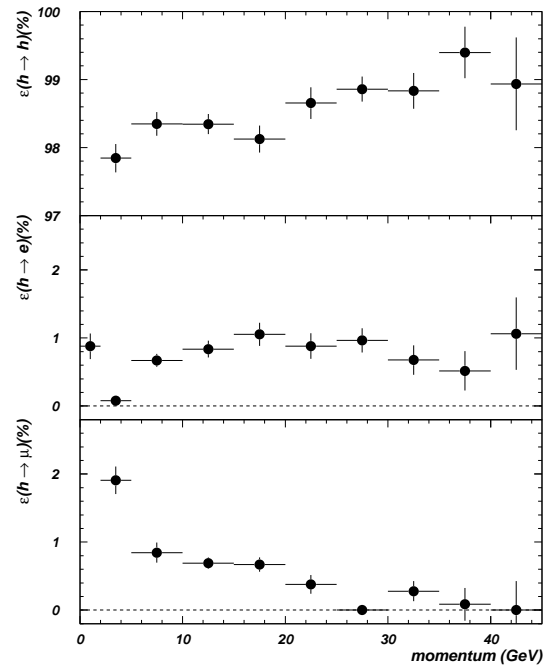,width=70mm}}
   \caption{Hadron identification efficiency and misidentification 
            probabilities obtained from the $\tau\tau$
            Monte Carlo, corrected from data using the control samples, 
            for the 1994-1995 data set. }
\label{ptidh}
\end{figure}

\begin{figure}[t]
   \centerline{\psfig{file=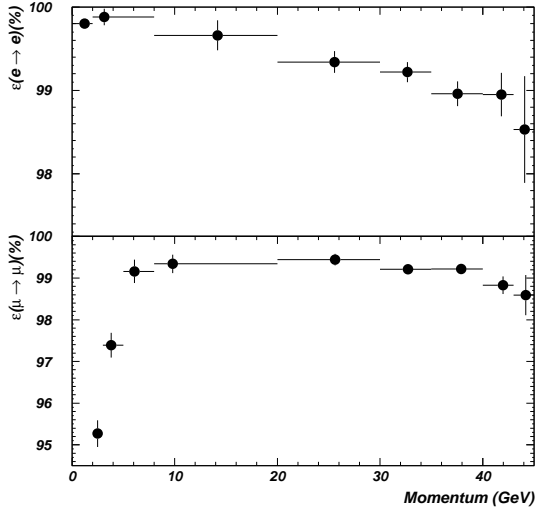,width=70mm}}
   \caption{Lepton identification efficiencies  obtained from the $\tau\tau$
            Monte Carlo, corrected from data using the control samples,
            for the 1994-1995 data set. }
\label{ptidl}
\end{figure}

The performance of the particle identification has been studied in detail 
using control samples of Bhabha events, $\mu\mu$ pairs, 
$\gamma\gamma$-induced lepton pairs and hadrons from $\pi^0$-tagged 
$\tau$ decays over the full angular and momentum range~\cite{aleph13_l}.
Measured efficiencies and misidentification probabilities are given in
Figs.~\ref{ptidh} and \ref{ptidl}.

\subsection{Photon identification}
\label{photid}

The high collimation of $\tau$ decays at LEP energies quite often makes
photon reconstruction difficult, since these photons are close to one
another or close to the showers generated by charged hadrons. Of particular
relevance is the rejection of fake photons which may occur because of
hadronic interactions, fluctuations of electromagnetic showers, or the
overlapping of several showers. These problems reach a tolerable level
thanks to the fine granularity of ECAL, in both transverse and 
longitudinal directions, but nevertheless they require the development 
of proper and reliable methods in order to correctly identify 
photon candidates.  

A cluster in ECAL is accepted as a photon candidate if its 
energy exceeds 300~MeV and if its barycentre is at least 2~cm away from 
the closest track extrapolation.

A likelihood method is used for discriminating between genuine 
and fake photons. For every cluster, an estimator $P_{\gamma}$ 
is defined, $P_{\gamma}=0,1$ corresponding to fake and real 
photons,respectively. It is constructed using probability 
densities obtained by simulation, but corrected through 
detailed comparisons between data and Monte Carlo.

Discriminating variables for each photon candidate used are the distance 
to the charged track, the distance 
to the nearest photon, a parameter from the clustering
process, the fractions of energy deposition in ECAL stacks 1, 
2 and 3, and a parameter related to the 
transverse energy distribution. 
Major improvements were introduced at this stage in the analysis compared 
to the previous one~\cite{aleph13_h}, mainly in the choice of variables
and in the use of energy-dependent reference distributions.

\begin{figure}[t]
   \vspace{9pt}
   \centerline{\psfig{file=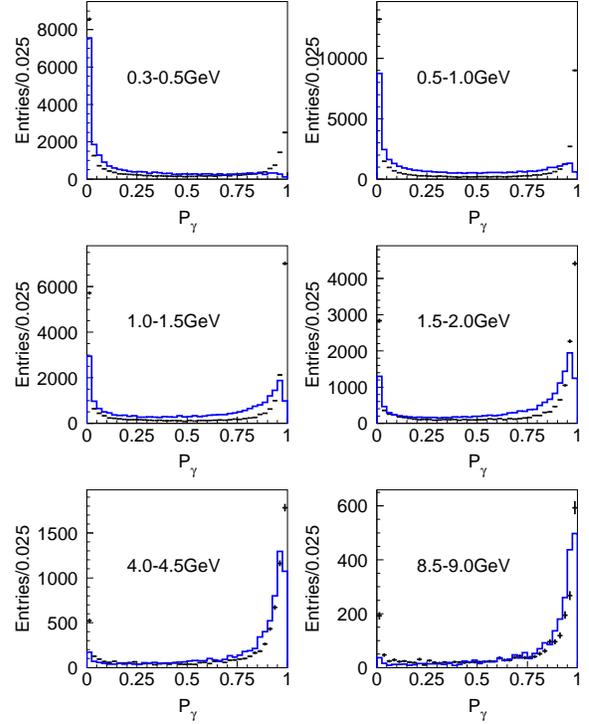,width=75mm}}
   \caption{Comparison of the probability distributions of a cluster to be 
   identified as a genuine photon before (histogram) and after (error bars)
   introducing energy-dependent reference distributions 
   and an improved set of discriminating variables.}
\label{prbg_old_new}
\end{figure}

Fig.~\ref{prbg_old_new} shows the comparison of the photon identification
probability ($P_{\gamma}$) before and after introducing energy-dependent
reference distributions and optimized variables. A spectacular improvement is
observed in genuine and fake photon discrimination: at low energy, a clear
contribution of genuine photons can be seen, while at high energy, 
a small, but well identified, fake photon component shows up.
 
Better photon energy calibration is also achieved compared to the previous 
analysis. The procedure, aiming at a relative calibration between data
and simulation, is implemented in several steps. First, a calibration 
is done using electrons from $\tau$ decays,
treating the ECAL information as for a neutral cluster. This method is 
reliable only above 2 GeV because of the electron track curvature in the
magnetic field. Then the final calibration is achieved 
by comparing the reconstructed $\pi^0$ mass and resolution as a function 
of energy.

Photon conversions are reconstructed following the previous 
analysis~\cite{aleph13_h}.

\subsection{$\pi^0$ reconstruction}
\label{pi0rec}

Three different kinds of $\piz$'s are reconstructed: resolved $\piz$
from two-photon pairing, unresolved $\piz$ from merged clusters, and residual
$\piz$ from the remaining single photons after removing radiative,
bremsstrahlung and fake photons with a likelihood method.

A $\piz$ estimator $D_{ij}$ is defined~\cite{aleph13_h}, taking into 
account the photon identification probabilities and the compatibility of the
pair with the $\piz$ mass hypothesis.

As the $\pi^0$ energy increases it becomes more difficult to resolve
the two photons and the clustering algorithm may yield a single cluster.
The two-dimensional energy distribution in the plane perpendicular to the
shower direction is examined and energy-weighted moments are computed. 
Assuming only two photons are present, the second moment provides a
measure of the $\gamma\gamma$ invariant mass~\cite{alephperf}. 
Clusters with mass greater than 0.1~GeV/$c^2$ are taken as unresolved 
$\piz$'s.

After the pairing and the cluster moment analysis, all the remaining 
photons inside a cone of $30^{\circ}$ around the jet axis are considered.
The radiative and bremsstrahlung photons are selected using the 
same method as described in the previous analysis~\cite{aleph13_h}.
Radiative and bremsstrahlung photons are not used
in the $\tau$ decay classification discussed below. 
The behaviour of these estimators was studied in Ref.~\cite{aleph13_h}.
The agreement between the number of bremsstrahlung photons in data and
simulation is good, but concerns mainly the electron $\tau$ decay channel
where this contribution is important (however it does not affect the
rate). Bremsstrahlung photons (i.e. radiation along the final state charged 
pions) in hadronic channels are at a much lower level and they are difficult 
to pick up unambiguously in the data. To estimate the effect of this 
contribution one therefore largely relies on the description of radiation at
the generator level in the Monte Carlo~\cite{photos}.

\subsection{Decay classification}
\label{class}

Each $\tau$ decay is classified topologically according to the number of 
hadronic charged tracks, their particle identification and the number of 
$\piz$'s reconstructed. While for 1-prong and 5-prong channels 
the exact multiplicity is required, the track number in 3-prong channels
is allowed to be 2, 3 or 4, in order to reduce systematic 
effects due to tracking and secondary interactions.

The definition of the leptonic channels requires an identified electron 
or muon with any number of photons. Some cuts on the total final state 
invariant mass are introduced to reduce feedthrough from hadronic 
modes~\cite{aleph13_l}.Also some decays with at least two good electron 
tracks are now included, when one or more of such tracks
are classified as converted photons.

In the previous analysis~\cite{aleph13_h}, the $\thd$,  $\tht$ and $\hq$ 
channels suffered from large backgrounds and consequently 
had a low signal-to-noise ratio. 
Most of these backgrounds are due to secondary interactions 
of the hadronic track with material in the inner detector part. 
In order to improve the definition of these channels the following steps 
are taken: (1) require the exact charged multiplicity $n_{ch}=3$ 
(instead of 2, 3, or 4) for $\thd$ and $\tht$, (2) demand a maximum 
impact parameter of charged tracks less than 0.2~cm (instead of 2~cm) 
for $\tht$, and (3) require the number of resolved $\pi^0$s in $\tht$ 
to be 3 or 2, and 4 or 3 in $\hq$.
With these tightened cuts the signal-to-noise ratio improves significantly 
with a small loss of efficiency.

It should be emphasized that all hemispheres from the
selected $\tau\tau$ event sample are classified, except for single tracks 
going into an ECAL crack (but for identified muons) or with a momentum less
than 2 GeV (but for identified electrons and hemispheres with at least
one reconstructed $\piz$). These latter decays, in addition to 
the rejected ones in the $\thd$ or $\tht$ channels are put in a special
class, labelled 14, which therefore collects all the non-selected 
hemispheres. By definition, the sample in class-14 does not correspond 
to one physical $\tau$ decay mode. In fact, it follows from 
the simulation that this class comprises about 21\% electron, 
27\% muon, 41\% 1-prong hadronic and 11\% 3-prong
$\tau$ decays. However, consideration of class-14 events can test if
the rejected fraction is correctly understood, as discussed later.

The numbers of $\tau$'s classified in each of the 
considered decay channels are listed in Table~\ref{nobs_dt}.

\begin{table}
\caption{Numbers of reconstructed and estimated non-$\tau$ background events
         in 1991-1993 and 1994-1995 data sets in the different
         considered topologies.}
\begin{center}
{
\small
\setlength{\tabcolsep}{0.42pc}
\begin{tabular}{l|rr|rr}
\hline\hline
class &   91-93&  &   94-95&   \\\hline
 \clsi  &      22405 & $598 \pm 46$ &  33100 & $745 \pm 58$   \\ 
 \clsii  &     22235 & $409 \pm 45$ &  32145 & $380 \pm 40$  \\ 
 \clsiii  &    15126 & $ 93 \pm 11$ &  22429 & $100 \pm 13$  \\ 
 \clsiv  &     32282 & $141 \pm 22$ &  49008 & $178 \pm 26$  \\ 
 \clsv  &      12907 & $ 44 \pm  9$ &  18317 & $ 81 \pm 16$  \\ 
 \clsvi  &      2681 & $ 26 \pm  7$ &   3411 & $ 35 \pm  9$  \\ 
 \clsxiii  &     458 & $ 12 \pm  3$ &    499 & $ 19 \pm  5$  \\ 
 \clsvii  &    11610 & $ 87 \pm 20$ &  17315 & $129 \pm 30$  \\ 
 \clsviii  &    6467 & $ 97 \pm 23$ &   9734 & $165 \pm 39$  \\ 
 \clsix  &      1091 & $ 27 \pm  7$ &   1460 & $ 36 \pm 10$  \\ 
 \clsx  &        124 & $ 13 \pm  4$ &    150 & $ 25 \pm  7$  \\ 
 \clsxi  &        60 & $  3 \pm  1$ &    105 & $  7 \pm  2$  \\ 
 \clsxii  &       36 & $ 16 \pm  5$ &     59 & $ 21 \pm  6$  \\ 
 14  &     4834 & $249 \pm 38$ &   7100 & $303 \pm 52$  \\ 
\hline
 sum      &   132316 & $1815 \pm 86$& 194832 & $2224 \pm 107$ \\
\hline\hline
\end{tabular}
}
\label{nobs_dt}
\end{center}
\end{table}

The KORAL07 generator~\cite{was} in the Monte Carlo simulation incorporates
all the decay modes considered in this analysis, except for 
the $h 4\pi^0$ decay channel. In the latter case, 
a separate generation was done where one of the produced 
$\tau$ is made to decay into that mode using a phase space model for the 
hadronic final state, while the other $\tau$ was treated
with the standard decay library. The complete behaviour 
between the generated decays and their reconstructed counterparts
using the decay classification is embodied in the efficiency matrix. This
matrix $\eff_{ji}$ gives the probability of a $\tau$ decay generated
in class $j$ to be reconstructed in class $i$. Obtained 
initially using the simulated samples, it is corrected for effects where
data and simulation can possibly differ, such as particle identification,
as discussed previously, and photon identification as affected by the
presence of fake photons.

\subsection{Adjusting the number of fake photons in the simulation}
\label{nbfake}

As could be expected, the number of fake photons in the simulation does 
not agree well with the rate in data. Indeed, it is observed that 
the rate of low-energy simulated fake photons is insufficient.
This effect will affect the classification of the reconstructed 
final states in the Monte Carlo and bias the efficiency matrix constructed 
from this sample. A procedure is developed to correct for this effect.

Taking explicitely into account the number $k$ of fake photons 
in a $\tau$ decay, the efficiency  matrix is taken as
$\eff_{jik}$, {\it i.e.} the efficiency for a produced event in class $j$ 
with $k$ fake photons reconstructed in class $i$. In this way the effect of
the different fake photon multiplicities in data and simulation can be 
explicitly taken into account, assuming that fake photons are randomly
produced. Then, it is sufficient to determine one factor $f_j$ in each 
produced class $j$ in order to quantify the deficit in the simulation.
These factors are obtained from fits of the observed $P_\gamma$ distributions
in data and the number of fake photons produced in the simulation.

In the published analysis using 1991-1993 data~\cite{aleph13_h} 
a much less sophisticated approach was taken. The deficit of fake photons
in the simulation was determined in a global way to be $(16\pm 8)\%$, 
common to all channels. Since it would have been delicate to generate
extra fake photons, the procedure chosen was to actually do the opposite,
{\it i.e.} randomly kill identified (in the sense of the matching
procedure discussed above) fake photons in the simulation, determine the
new efficiency matrix and take as corrections the sign-reversed deviations.
   
It is clear that the current way of dealing with the fake photon problem 
is both more precise and more reliable. The fact that different channels
are treated separately provides a handle on the different origins of the 
fake photons, since, for example, fake photons in the \clsiii ~class
only originate from hadronic interactions, whereas they come from both
hadronic interactions and photon shower fluctuations in the  
\clsiv ~class.

\section{Results}

\subsection{Determination of the branching ratios}
\label{brdet}

The branching ratios are determined using
\beqn
 n^{obs}_i - n^{bkg}_i &=& \sum_{j} \eff_{ji} N^{prod}_j\\
 B_j &=& \frac {N^{prod}_j} {\sum_{j} N^{prod}_j}
\eeqn
where
$n^{obs}_i$ is the observed events number in reconstructed class $i$,
$n^{bkg}_i$ the non-$\tau$ background in reconstructed class $i$,
$\eff_{ji}$ the efficiency of a produced class $j$ event reconstructed
as class $i$, and
$N^{prod}_j$ the produced events number of class $j$. The class numbers
$i$, $j$ run from 1 to 14, the last one corresponding to the rejected
$\tau$ candidates.

The efficiency matrix $\eff_{ji}$ is determined from the Monte Carlo, 
but corrected using data for many effects such as particle identification
efficiency, $\tau\tau$ selection efficiency and fake photon simulation. 

The analysis assumes a standard $\tau$ decay description. One could
imagine unknown decay modes not included in the simulation, but since
large detection efficiencies are achieved in the $\tau\tau$ 
selection which is therefore robust, these decays would be
difficult to pass unnoticed. An independent measurement of the
branching ratio for undetected decay modes, using a direct search with
a one-sided $\tau$ tag, was done in ALEPH~\cite{aleph_undetect},
limiting this branching ratio to less than 0.11\% at 95\% CL. This
result justifies the assumption that the sum of the branching ratios 
for visible $\tau$ decays is equal to one.

The branching ratios are obtained and listed 
in Table~\ref{br_12345}.

\begin{table*}
\caption{Branching ratios (\%) for the reconstructed topologies
         (quasi-exclusive modes) from 1991-1993 and 1994-1995 data sets; 
         the first error is statistical and the second is systematic.}
\begin{center}
\setlength{\tabcolsep}{3.0pc}
\begin{tabular}{|l|c|c|}
\hline\hline
class    &  91-93  & 94-95  \\\hline
 \clsi  &   17.859 $\pm$  0.112 $\pm$  0.058 &   17.799 $\pm$  0.093 $\pm$  0.045 \\ 
 \clsii  &   17.356 $\pm$  0.107 $\pm$  0.055 &   17.273 $\pm$  0.087 $\pm$  0.039 \\ 
 \clsiii  &   12.238 $\pm$  0.105 $\pm$  0.104 &   12.058 $\pm$  0.088 $\pm$  0.083 \\ 
 \clsiv  &   26.132 $\pm$  0.150 $\pm$  0.104 &   26.325 $\pm$  0.123 $\pm$  0.090 \\ 
 \clsv  &    9.680 $\pm$  0.139 $\pm$  0.124 &    9.663 $\pm$  0.107 $\pm$  0.105 \\ 
 \clsvi  &    1.128 $\pm$  0.110 $\pm$  0.086 &    1.229 $\pm$  0.089 $\pm$  0.068 \\ 
 \clsxiii  &    0.227 $\pm$  0.056 $\pm$  0.047 &    0.163 $\pm$  0.050 $\pm$  0.040 \\ 
 \clsvii  &    9.931 $\pm$  0.097 $\pm$  0.072 &    9.769 $\pm$  0.080 $\pm$  0.059 \\ 
 \clsviii  &    4.777 $\pm$  0.093 $\pm$  0.074 &    4.965 $\pm$  0.077 $\pm$  0.066 \\ 
 \clsix  &    0.517 $\pm$  0.063 $\pm$  0.050 &    0.551 $\pm$  0.050 $\pm$  0.038 \\ 
 \clsx  &    0.016 $\pm$  0.029 $\pm$  0.020 &   -0.021 $\pm$  0.023 $\pm$  0.019 \\ 
 \clsxi  &    0.098 $\pm$  0.014 $\pm$  0.006 &    0.098 $\pm$  0.011 $\pm$  0.004 \\ 
 \clsxii  &    0.022 $\pm$  0.010 $\pm$  0.009 &    0.028 $\pm$  0.008 $\pm$  0.007 \\ 
 14  &    0.017 $\pm$  0.043 $\pm$  0.042 &    0.099 $\pm$  0.035 $\pm$  0.037 \\ 
\hline\hline
\end{tabular}
\label{br_12345}
\normalsize
\end{center}
\end{table*}

\subsection{Determination of systematic uncertainties}
\label{system}

Wherever possible the efficiencies relevant to the analysis have been 
determined using ALEPH data, either directly on the $\tau\tau$ sample
itself or on specifically selected control samples, as for example 
in the case of particle identification. The resulting efficiencies 
are thus measured with known statistical errors. 

In some cases the procedure is less straightforward and involves a model 
for the systematic effect to be evaluated. An important example is given
by the systematics in the simulation of fake photons in ECAL. 
In such cases the evaluation of the systematic error not only takes into
account the statistical aspect, but also some estimate of the systematics
involved in the assumed model. The latter is obtained from studies where
the relevant parameters are varied in a range consistent with the
comparison between data and Monte Carlo distributions.

Quite often a specific cut on a given variable can be directly studied.
The comparison between the corresponding distributions, respectively
in data and Monte Carlo, provides an estimate of a possible discrepancy 
whose effect would be to change the assumed efficiency of the cut. If a
significant deviation is observed, the correction is applied to the
simulation to obtain the nominal branching ratio results, while the error
on the deviation provides the input to the evaluation of the systematic
uncertainty. The analysis is therefore repeated with a full 
re-classification of all the measured $\tau$ decay candidates, changing
the incriminated variable by one standard deviation. Since the new samples
slighly differ from the nominal ones because of feedthrough between the
different channels, the modified results are affected both by the 
systematic change in the variable value and the statistical fluctuation
from the event sample which is uncommon to both samples. In this case
the final systematic uncertainty is obtained by adding linearly the modulus
of the systematic deviation observed and the statistical error from
the monitored uncommon sample. This procedure is followed 
for all the systematic studies.

Finally, the systematic deviations for each study are obtained with
their sign in each measured decay channel, thus providing the full
information on the correlations between the results and allowing the 
corresponding covariance matrix to be constructed.

The most important systematic uncertainties originate from photon 
identification and $\piz$ reconstruction, and secondary interactions.

\subsection{From reconstructed classes to exclusive modes}

So far branching fractions have been determined in 13 classes
corresponding to major $\tau$ decay modes. However, these classes 
still contain the contributions
from final states involving kaons. The latter are coming from 
Cabibbo-suppressed $\tau$ decays or modes with a $K\overline{K}$ pair,
both characterized by small branching ratios compared to the nonstrange
modes without kaons.
Complete analyses of $\tau$ decays involving neutral 
or charged kaons have been performed by ALEPH on the full LEP I 
data~\cite{alephk3,alephks,alephkl}. They are summarized and
measurements with $K^0_S$ and $K^0_L$ are combined in Ref.~\cite{alephksum}.

The $\tau$ decays involving $\eta$ or $\omega$ mesons also require 
special attention in this analysis because of their electromagnetic
decay modes. Indeed the final state classification relies in part 
on the $\piz$ multiplicity, thereby assuming that all photons ---but
those specifically identified as bremsstrahlung or radiative---
originate from $\piz$ decays. Therefore the non-$\piz$ photons from
$\eta$ and $\omega$ decays are treated as $\piz$ candidates in the
analysis and the systematic bias introduced by this effect must be
evaluated. The corrections are based on specific measurements by ALEPH 
of $\tau$ decay modes containing those mesons~\cite{alepheta}. Thus
the final results correspond to exclusive branching ratios obtained
from the values measured in the topological classification, 
corrected by the removed contributions from $K$, $\eta$ and $\omega$  
modes measured separately, taking into account through the Monte Carlo
their specific selection and reconstruction efficiencies to enter the
classification. This bookkeeping takes into account all the
major decay modes of the considered mesons~\cite{pdg2000}, including
the isospin-violating $\omega \rightarrow \pi^+\pi^-$ decay mode.
The main decay modes into consideration are $\pi \omega$, $\pi \pi^0 \omega$
and $\pi \pi^0 \eta$ with branching fractions of $(2.26 \pm 0.18)~10^{-2}$,
$(4.3 \pm 0.5)~10^{-3}$, and $(1.80 \pm 0.45)~10^{-3}$~\cite{alepheta}, 
respectively. The first two numbers are derived from 
the branching ratios for the $3\pi \pi^0$ and $3\pi 2\pi^0$ 
modes obtained in this analysis and the measured $\omega$ 
fractions of $0.431 \pm 0.033$ from ALEPH~\cite{alepheta} 
and the average value, $0.78 \pm 0.06$, from ALEPH~\cite{alepheta} 
and CLEO~\cite{cleoomega}, respectively.

Some much smaller contributions with $\eta$ and $\omega$ have been 
identified and measured by CLEO~\cite{cleoeta3pi} with the decay modes 
$\tau \rightarrow \eta\pi^-\pi^+\pi^-$ ($(2.4 \pm 0.5)~10^{-4}$), 
$\tau \rightarrow \eta\pi^-2\piz$ ($(1.5 \pm 0.5)~10^{-4}$),
$\tau \rightarrow \omega\pi^-\pi^+\pi^-$ ($(1.2 \pm 0.2)~10^{-4}$),
and $\tau \rightarrow \omega\pi^-2\piz$ ($(1.5 \pm 0.5)~10^{-4}$).
Even though the corrections from these channels are very small they have
been included for the sake of completeness. Finally, another very small
correction has been applied to take into account the $a_1$ 
radiative decay into $\pi \gamma$ with a branching fraction of 
$(2.1 \pm 0.8)~10^{-3}$ obtained from Ref.~\cite{zielinski}.

\subsection{Overall consistency test}

Rejected $\tau$ hemispheres because of charged particle identification
cuts (2 GeV minimum momentum and ECAL crack veto for some 1-prong modes, 
strict definition of higher multiplicity channels) are placed in class 14.
As already emphasized, this sample does not correspond to a nominal
$\tau$ decay mode and should be explained by all other measured fractions
in the other classes and the efficiency matrix. Thus the determination of
a hypothetical signal in this class is a measure of the level of 
consistency achieved in the analysis.

For this determination the efficiency of the possible signal in class 14
is taken to be 100\%. The results, already shown in Table~\ref{br_12345}
separately for the 1991-1993 and 1994-1995 data sets, are consistent 
and are combined to give $B_{14}= (0.066 \pm (0.027)_{stat} 
\pm (0.021)_{syst,c} \pm (0.025)_{syst,unc})\%$, where the last two errors
refer to the common and uncommon uncertainties from the two data sets.
With a combined error of $0.042\%$ this value is consistent with zero
and provides a good check of the overall procedure
at the nontrivial 0.1\% level for branching ratios. It is interesting to
note that this value coincides, approximately and accidentally, with
the limit achieved of 0.11\% at 95\% CL in a direct search for 
'invisible' decays not selected in the 13-channel classification.

In the following it is assumed that all $\tau$ decay modes 
have been properly considered at the 0.1\% precision level and
no physics contribution beyond standard $\tau$ decays is further allowed.
Thus the quantity $B_{14}$ is now constrained to be zero.

It can be further noticed that this analysis provides a branching ratio 
in the $3\pi 3\piz$ class which is consistent with zero for both
1991-1993 and 1994-1995 data sets (see Table~\ref{br_12345}).
The result is therefore given as an upper limit at 95\% CL
\beq
 B_{3\pi 3\piz} < 4.9~10^{-4}
\eeq
consistent with the measurement made by CLEO~\cite{cleo6pi} yielding
$B_{3\pi 3\piz} = (2.2 \pm 0.5)~10^{-4}$. The final state is dominated by
$\eta$ and $\omega$ resonances~\cite{cleo6pi} and using other channels allows 
one to obtain a lower limit for this branching ratio, $(2.6 \pm 0.4)~10^{-4}$. 
In the following a value of $(3 \pm 1)~10^{-4}$ is used as input for this 
channel and the global analysis is performed in terms of the remaining 
12 defined channels which are refitted.

\begin{figure}[t]
   \centerline{\psfig{file=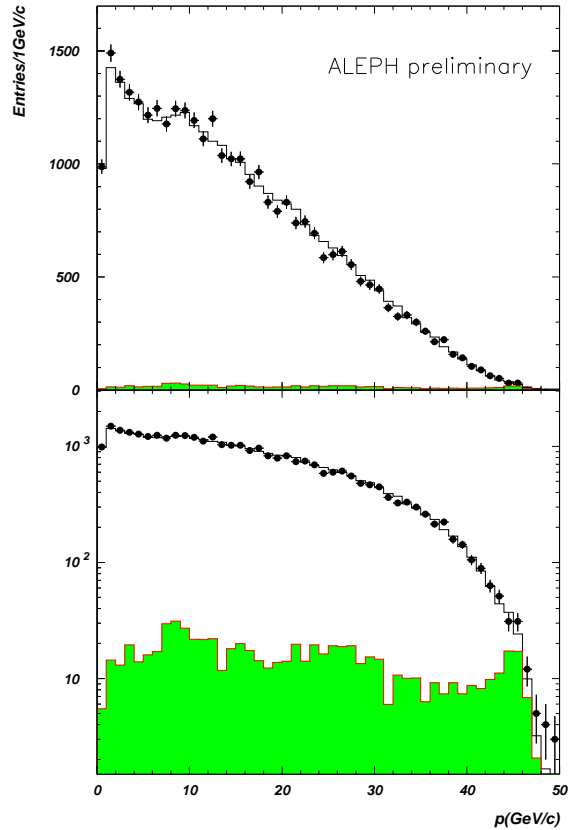,width=74mm}}
   \caption{Momentum spectrum in the electron channel in data (error bars) 
   and Monte Carlo (histogram) for the 1994-1995 data set. 
   The shaded histogram is the contribution of non-$\tau$ background.}
\label{spec_e}
\end{figure}

\begin{figure}[t]
   \centerline{\psfig{file=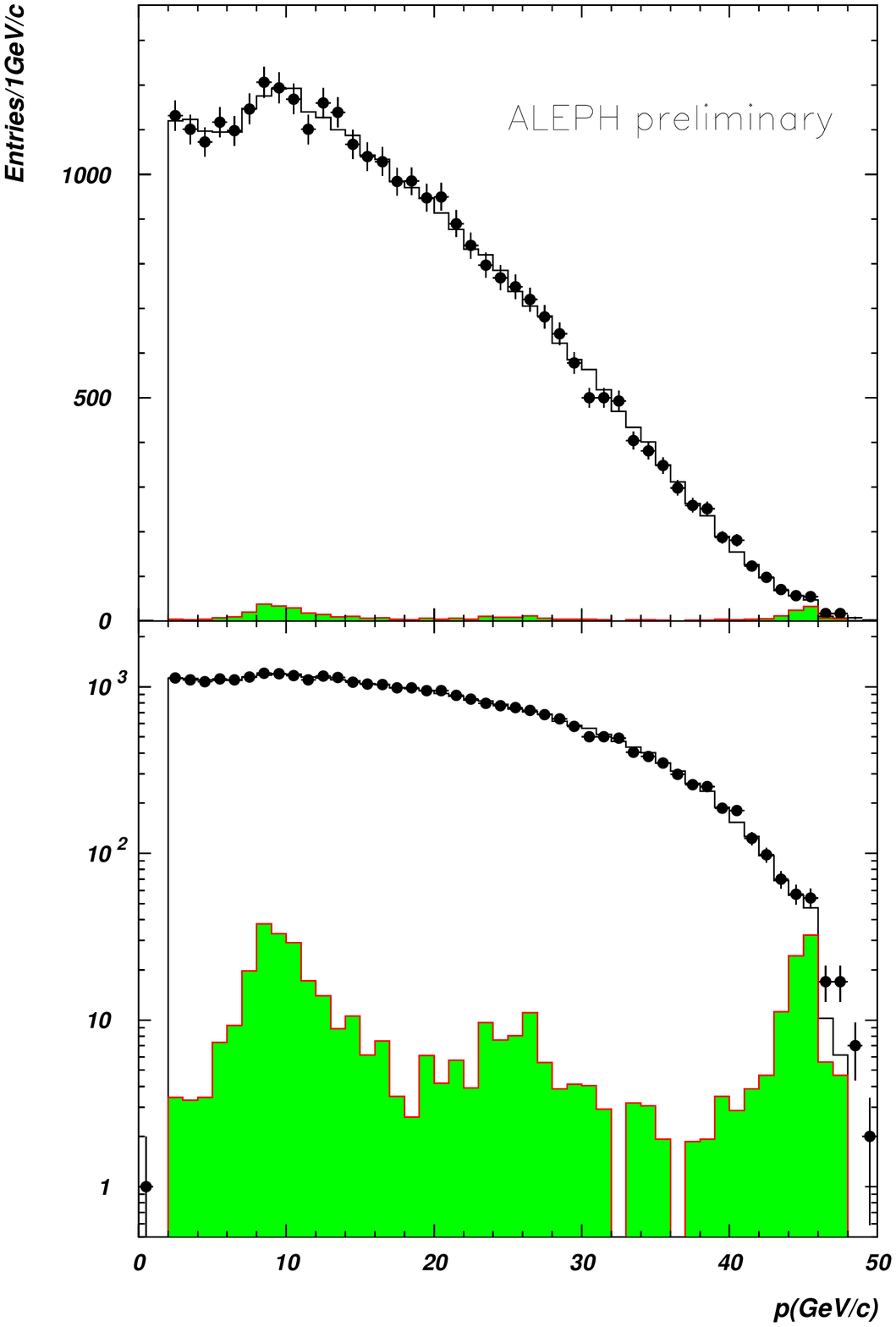,width=74mm}}
   \caption{Momentum spectrum in the muon channel in data (error bars) 
   and Monte Carlo (histogram) for the 1994-1995 data set. 
   The shaded histogram is the contribution of non-$\tau$ background.}
\label{spec_mu}
\end{figure}

\begin{figure}[t]
   \centerline{\psfig{file=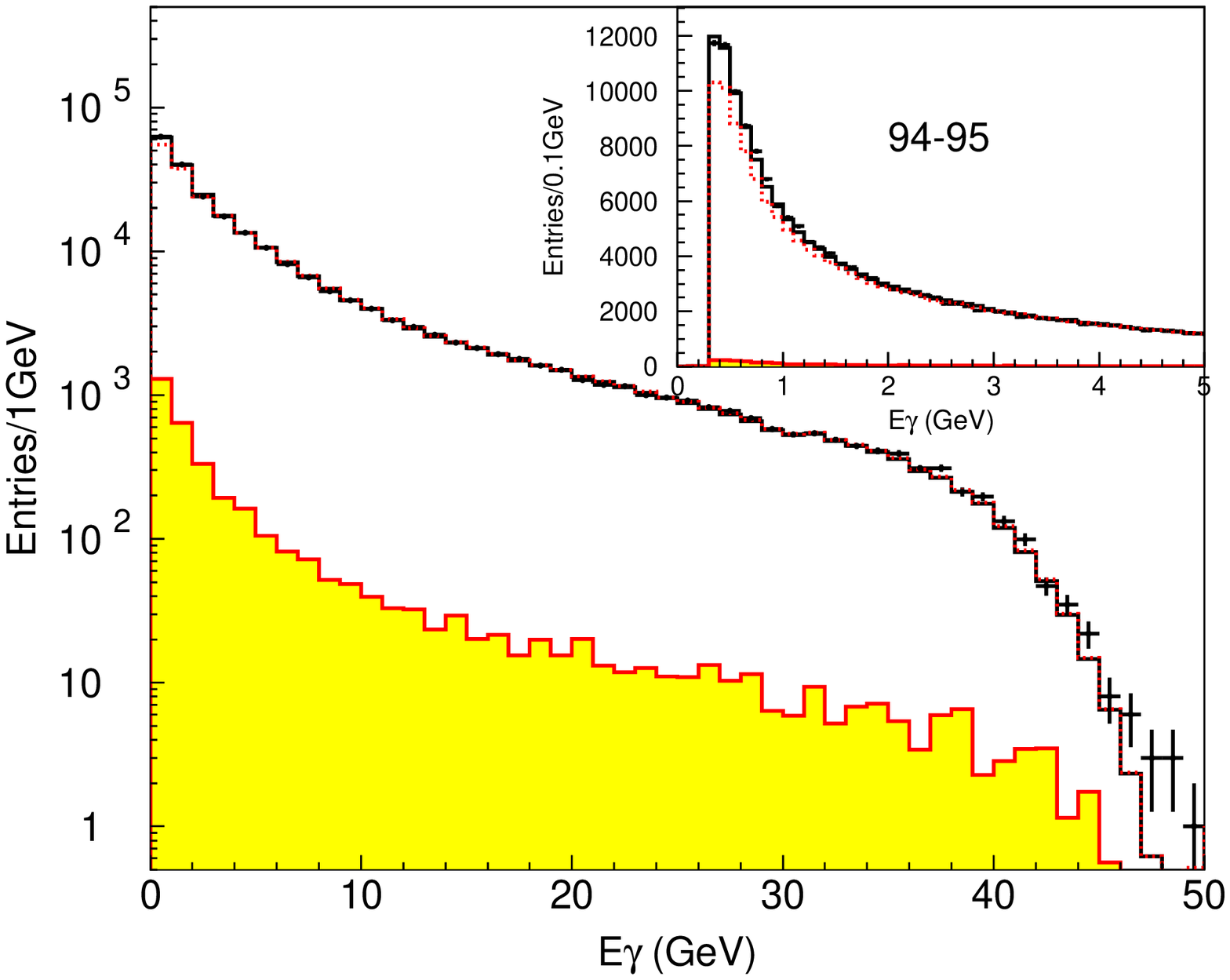,width=74mm}}
   \centerline{\psfig{file=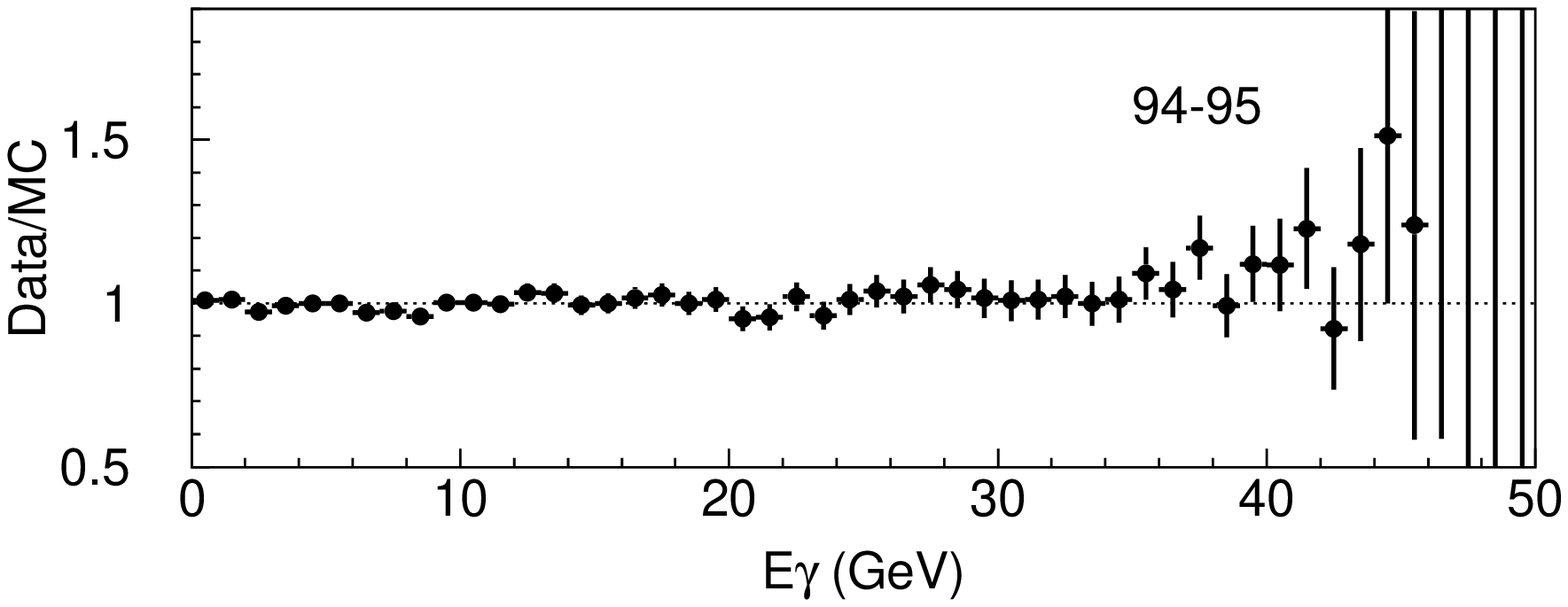,width=74mm}}
   \caption{Comparison of photon energy spectra for all $\tau$
   events in data (error bars) and Monte Carlo (histogram) after
   all the corrections for 1994-1995 data set. The dotted line shows the
   Monte Carlo distribution before correction and the shaded histogram is
   the contribution of non-$\tau$ background.}
\label{spec_gam}
\end{figure}

\begin{figure}[t]
   \centerline{\psfig{file=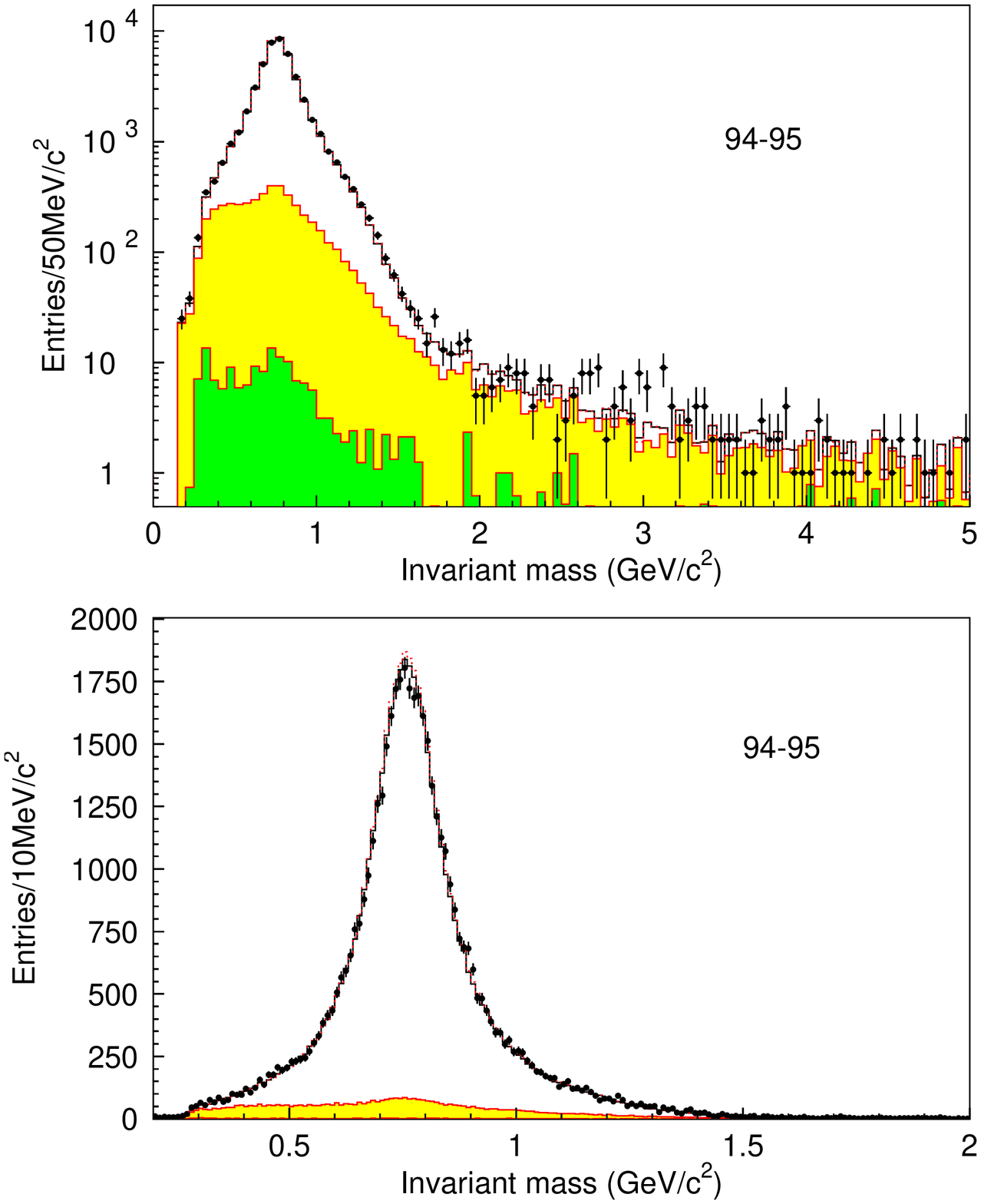,width=74mm}}
   \caption{Comparison of mass spectra in data (error bars)
   and Monte Carlo (histogram) after fake photon correction for the
   \clsiv sample in 1994-1995 data sample. The dark shaded histogram is
   the contribution of non-$\tau$ background and the light shaded
   is from $\tau$ feedthrough.}
\label{spec_mass}
\end{figure}

Figs.~\ref{spec_e}, \ref{spec_mu}, \ref{spec_gam} and \ref{spec_mass}
present a few comparisons between data and simulation, providing 
global checks on the data quality.

\subsection{Comparison of 1991-1993 and 1994-1995 results}

Since the same procedure is applied for the analyses of 1991-1993 
and 1994-1995 data the results must be consistent within the statistical 
errors of data and Monte Carlo. Good agreement is observed with a $\chi^2$
of 8.6 for 12 DF. In conclusion, the two independent data and Monte Carlo 
samples give consistent results.

\subsection{Final combined results}

Finally the two sets of results are combined. Whether one uses only 
statistical or total weights ---in the latter taking into account correlated
errors from dynamics and secondary interactions--- gives almost identical 
results. Using the total weights one obtains the final results shown in
Table~\ref{finalBR}. 

\begin{table}[t]
\caption{Combined results for the exclusive branching ratios (B) 
         for modes without kaons. The contributions from channels with 
         $\eta$ and $\omega$ are given separately, the latter only for the 
         electromagnetic $\omega$ decays. All results are from this analysis, 
         unless explicitly stated. The values labelled * and **,*** are 
         taken from ALEPH~\cite{alepheta} and 
         CLEO~\cite{cleoeta3pi,cleoomega}, respectively.}
\begin{center}
\small
\setlength{\tabcolsep}{0.17pc}
\begin{tabular}{lr}
\hline\hline
 mode & B $\pm\sigma_{\hbox{stat}} \pm \sigma_{\hbox{syst}}$
  [\%] \\\hline
 \phyi &    17.837 $\pm$  0.072 $\pm$  0.036 \\
 \phyii &    17.319 $\pm$  0.070 $\pm$  0.032 \\
 \phyiii &    10.828 $\pm$  0.070 $\pm$  0.078 \\
 \phyiv &    25.471 $\pm$  0.097 $\pm$  0.085 \\
 \phyv &     9.239 $\pm$  0.086 $\pm$  0.090 \\
 \phyvi &      0.977 $\pm$  0.069 $\pm$  0.058 \\
 \phyxiii &      0.112 $\pm$  0.037 $\pm$  0.035 \\
 \phyvii &     9.041 $\pm$  0.060 $\pm$  0.076 \\
 \phyviii &     4.590 $\pm$  0.057 $\pm$  0.064 \\
 \phyix &      0.392 $\pm$  0.030 $\pm$  0.035 \\
 \phyx  ~(estim.)&      0.013 $\pm$  0.000 $\pm$  0.010 \\
 \phyxi &      0.072 $\pm$  0.009 $\pm$  0.012 \\
 \phyxii &      0.014 $\pm$  0.007 $\pm$  0.006 \\
$\pi^- \pi^0 \eta^*$ & 0.180 $\pm$ 0.040 $\pm$ 0.020 \\
$\pi^- 2\pi^0 \eta^{**}$ & 0.015 $\pm$ 0.004 $\pm$ 0.003 \\
$\pi^- \pi^- \pi^+ \eta^{**}$ & 0.024 $\pm$ 0.003 $\pm$ 0.004 \\
$a_1^- (\rightarrow \pi^- \gamma)$~(estim.) & 0.040 $\pm$ 0.000 $\pm$ 0.020 \\
$\pi^- \omega (\rightarrow \pi^0 \gamma, \pi^+ \pi^-)^*$ & 0.253 $\pm$ 0.005 $\pm$ 0.017 \\
$\pi^- \pi^0 \omega (\rightarrow \pi^0 \gamma, \pi^+ \pi^-)^{*,***}$ & 0.048 $\pm$ 0.006 $\pm$ 0.007\\
$\pi^- 2\pi^0 \omega (\rightarrow \pi^0 \gamma, \pi^+ \pi^-)^{**}$ & 0.002 $\pm$ 0.001 $\pm$ 0.001 \\
$\pi^- \pi^- \pi^+\omega (\rightarrow \pi^0 \gamma, \pi^+ \pi^-)^{**}$ & 0.001 $\pm$ 0.001 $\pm$ 0.001 \\
\hline\hline
\end{tabular}
\normalsize
\label{finalBR}
\end{center}
\end{table}

The branching ratios obtained for the different channels are
correlated with each other. On one hand the statistical fluctuations
in the data and the Monte Carlo sample are driven by the multinomial
distribution of the corresponding events, producing well-understood
correlations . On the other hand the systematic effects also induce
important and nontrivial correlations between the different channels.
All the systematic studies were done keeping track of the correlated
variations in the final branching ratio results, thus allowing a proper
propagation of errors.
 
The present results are consistent with those of the previously published
ALEPH analysis~\cite{aleph13_l,aleph13_h}. The leptonic branching ratios
also agree within errors with the results of an independent ALEPH analysis
which does not rely on a global method~\cite{alephl_moriond}.

\section{Discussion of the results}
\label{phys}

\subsection{Comparison with other experiments}

\begin{figure}[t]
   \centerline{\psfig{file=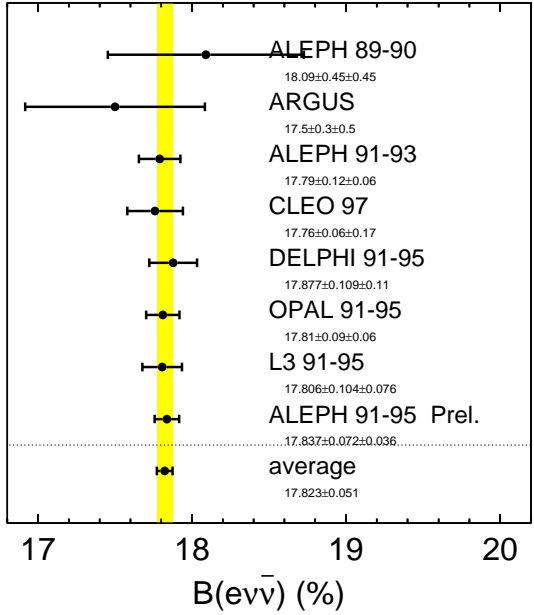,width=70mm}}
   \caption{Comparison of ALEPH measurement with published precise results
   from other experiments for $\tau \ra e \nu \bar{\nu}$.}
   \label{bre}
\end{figure}

\begin{figure}[t]
   \centerline{\psfig{file=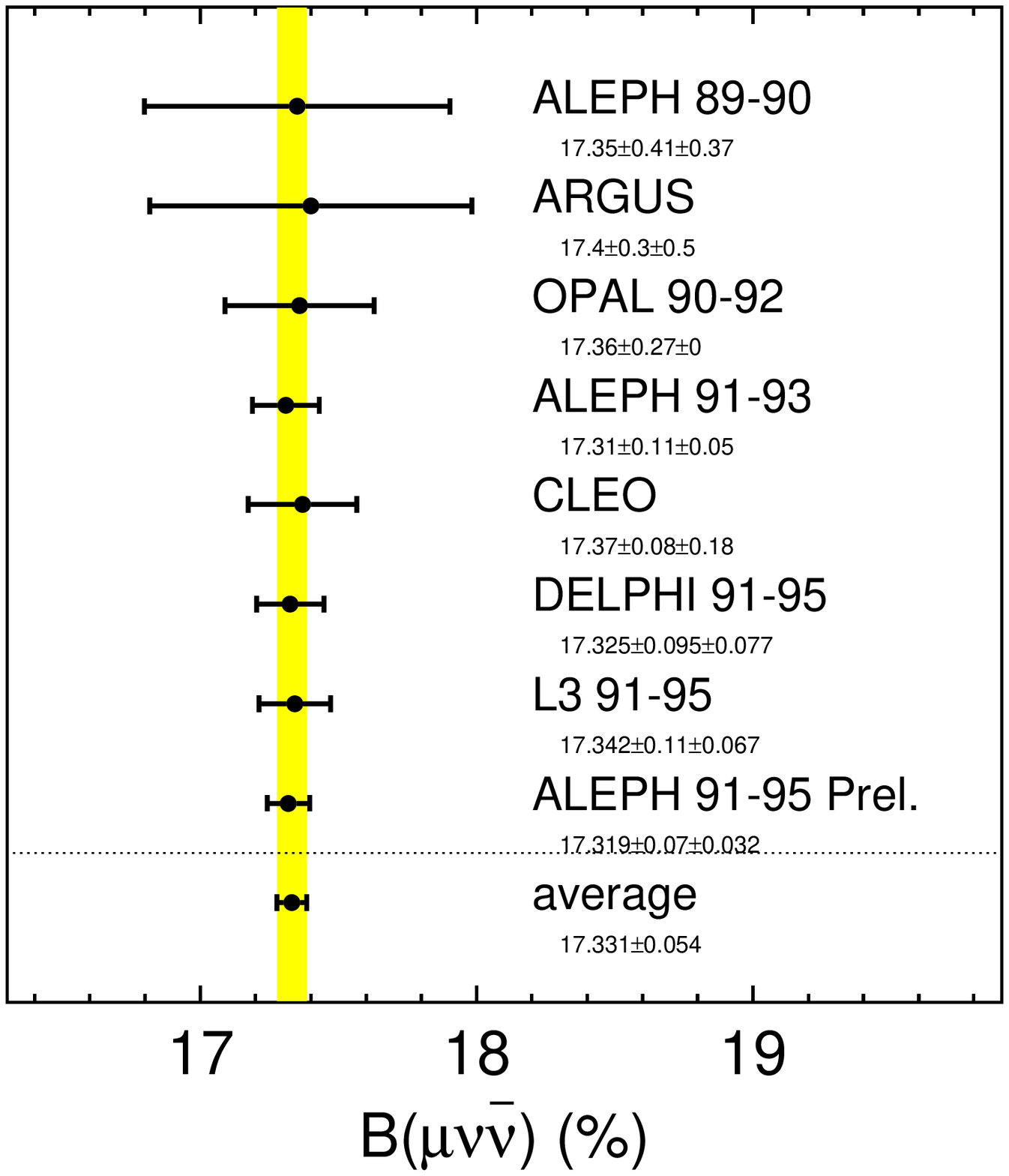,width=70mm}}
   \caption{Comparison of ALEPH measurement with published precise results
   from other experiments for $\tau \ra \mu \nu \bar{\nu}$.}
   \label{brmu}
\end{figure}

\begin{figure}[t]
   \centerline{\psfig{file=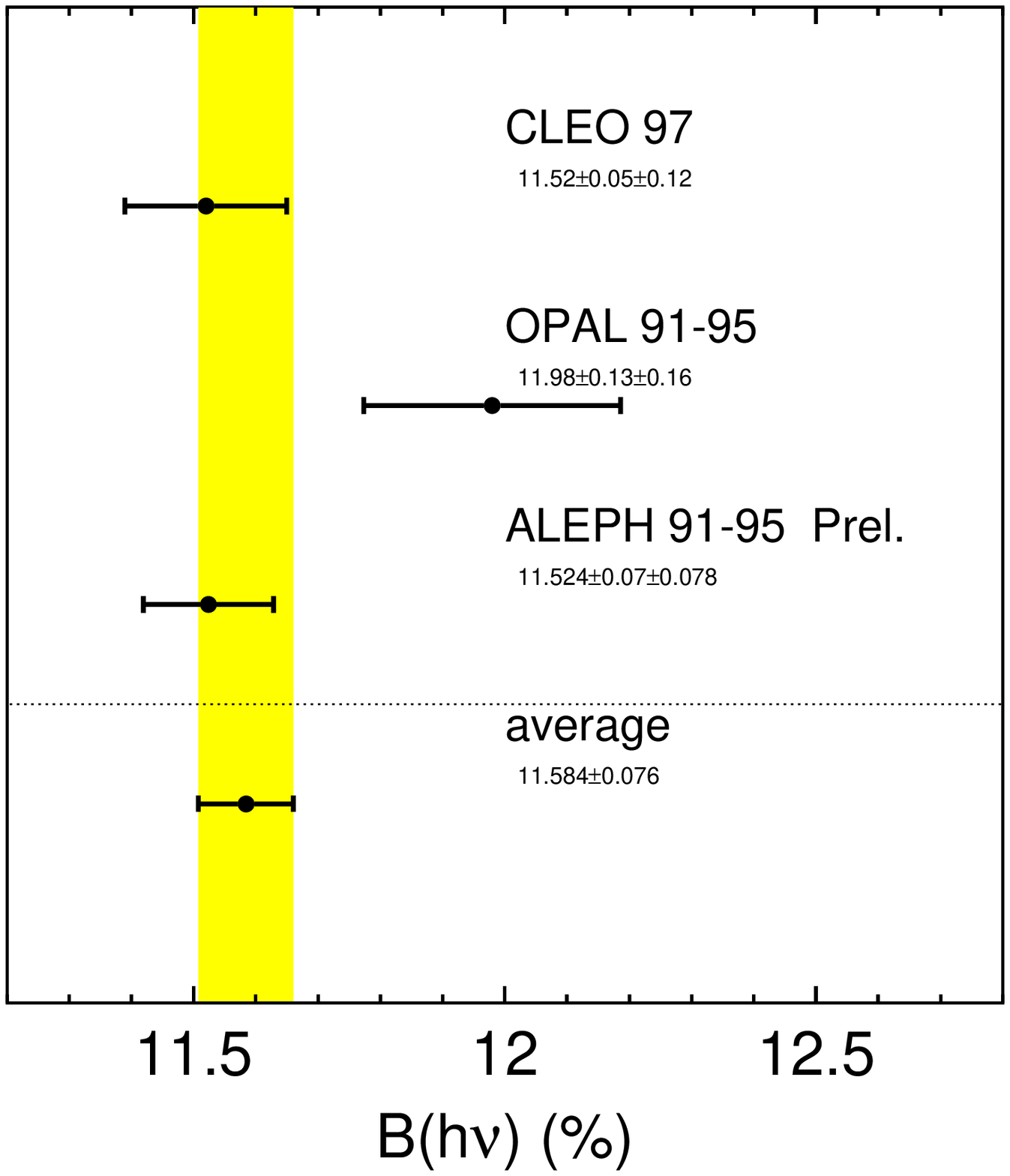,width=70mm}}
   \caption{Comparison of ALEPH measurement with published precise results
   from other experiments for $\tau \ra h \nu$ (sum of $\pi \nu$ and
   $K \nu$).}
   \label{brh}
\end{figure}

\begin{figure}[t]
   \centerline{\psfig{file=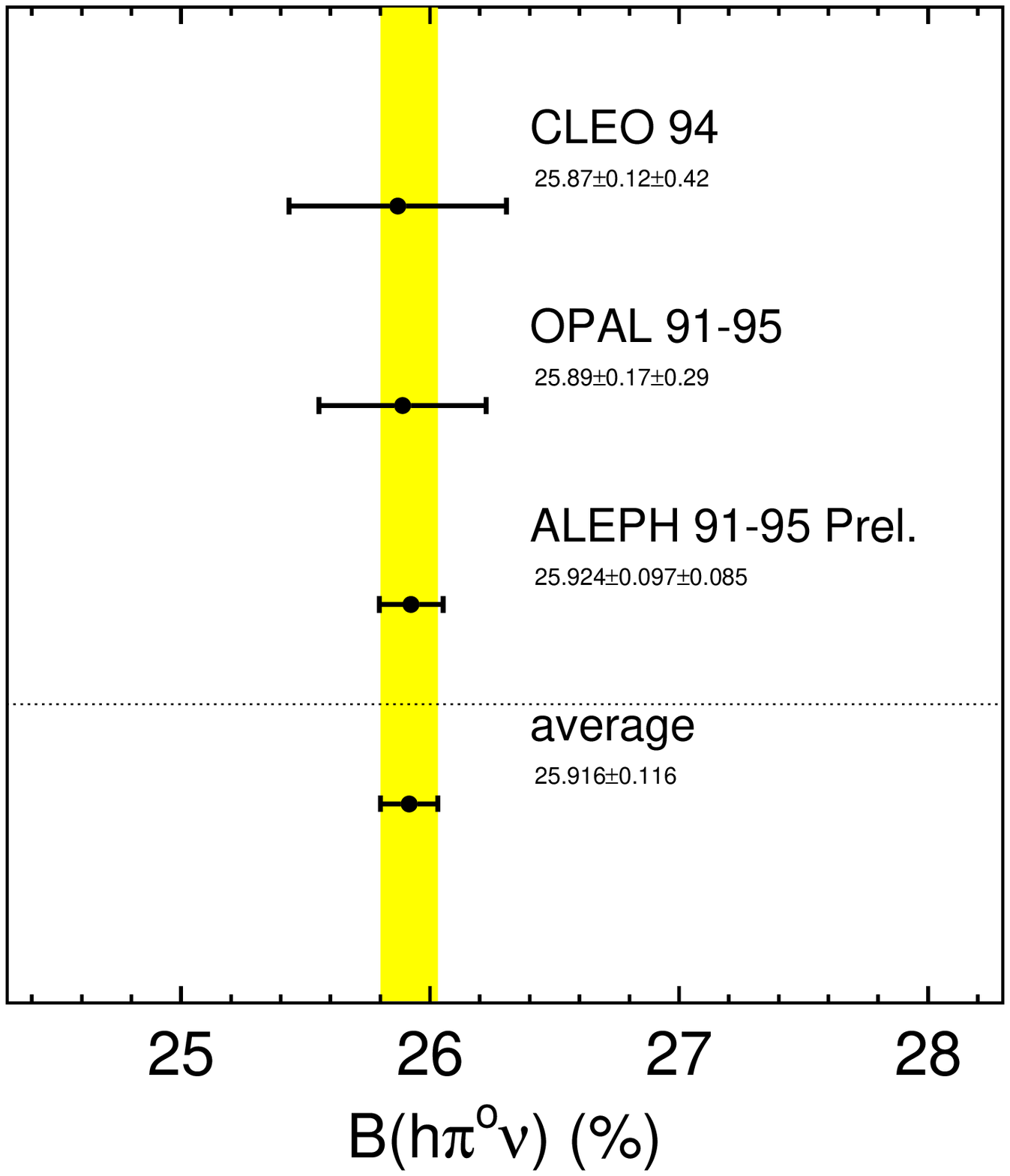,width=70mm}}
   \caption{Comparison of ALEPH measurement with published precise results
   from other experiments for $\tau \ra h\piz \nu $ (sum of $\pi \piz \nu$ and
   $K \piz \nu$).}
   \label{brhpi0}
\end{figure}

Figs.~\ref{bre}, \ref{brmu}, \ref{brh}, \ref{brhpi0} 
show that the results of this analysis are in good agreement 
with those from other most precise experiments. 
In all these cases, ALEPH achieves the best precision.

A meaningful comparison can be performed between the exclusive fractions 
and the topological branching ratios. Even though the latter have 
essentially no physics interest, their determination can constitute 
a valuable cross check as they depend only on selection efficiency, 
tracking, handling of secondary interactions and electron identification 
for photon conversions, and not on photon identification.
The results from this analysis can be compared in this way with a dedicated 
analysis recently performed by DELPHI~\cite{delphitopol}. 
Summing up appropriately (both analyses assume a negligible
contribution from hadronic multiplicities higher than 5, in agreement
with the 90\% CL limit by CLEO~\cite{cleo7pr} one gets 
$B_7 < 2.4~ 10^{-6}$)
\beqn
 B_3 &=& (14.652 \pm 0.067 \pm 0.086) ~\% \\
 B_5 &=& (0.093 \pm 0.009 \pm 0.012) ~\% 
\eeqn
in good agreement with the DELPHI values,
$B_3 = (14.569 \pm 0.093 \pm 0.048) ~\%$ and 
$B_5 = (0.115 \pm 0.013 \pm 0.006) ~\%$.
The rather small systematic uncertainty in the DELPHI results reflects 
the fact that a sharper study of hadronic interactions can be
performed when only charged particles are considered in the analysis.
In addition the modes with $K^0_s \rightarrow \pi^+ \pi^-$ decays are
subtracted statistically here, rather than trying to identify them on an 
event-by-event basis.

\subsection{Universality in the leptonic charged current}
\label{uni}

\subsubsection{$\mu$-e universality from the leptonic branching ratios}

In the standard V-A theory with leptonic coupling $g_l$ at the
$W l \overline{\nu}_l$ vertex, the $\tau$ leptonic partial width can be 
computed, including radiative corrections~\cite{marciano-sirlin} and
safely neglecting neutrino masses:
\beq
\Gamma(\tau \rightarrow \nu_\tau l \overline{\nu}_l (\gamma)) =
 \frac {G_\tau G_l m^5_\tau}{192 \pi^3} f\left(\frac {m^2_l}{m^2_\tau}\right)
 \delta^\tau_W \delta^\tau_\gamma~,
\eeq
where
\beqn
 G_l &=& \frac {g^2_l}{4 \sqrt{2} M^2_W}  \nonumber \\
 \delta^\tau_W &=& 1 + \frac {3}{5} \frac {m^2_\tau}{M^2_W} \nonumber \\
 \delta^\tau_\gamma &=& 1+\frac {\alpha(m_\tau)}{2\pi}
 \left(\frac {25}{4}-\pi^2\right) \nonumber \\
 f(x) &=& 1 -8x +8x^3 -x^4 -12x^2 {\rm ln}x 
\eeqn
Numerically, the W propagator correction and the radiative corrections
are small:
$\delta^\tau_W =1+2.9\cdot 10^{-4}$ and 
$\delta^\tau_\gamma=1+43.2\cdot 10^{-4}$.

Taking the ratio of the two leptonic branching fractions, a direct test
of $\mu$-e universality is obtained. The measured ratio
\beq
  \frac {B_\mu} {B_e} = 0.9709 \pm 0.0060 \pm 0.0029
\eeq
agrees with the expectation 
which is equal to 0.97257 when universality holds. Alternatively
the measurements yield the ratio of couplings
\beq
  \frac {g_\mu} {g_e} = 0.9991 \pm 0.0033 
\eeq
which is consistent with one.

This result is in agreement with the best test of $\mu$-e universality
of the W couplings obtained in the comparison of the rates for
$\pi \rightarrow \mu \overline{\nu}_\mu$ and 
$\pi \rightarrow e \overline{\nu}_e$ decays, where the results 
of the two most accurate experiments~\cite{triumf,psi} can be averaged
to yield $\frac {g_\mu} {g_e} = 1.0012 \pm 0.0016 $. The results have
comparable precision, but it should be pointed out that they are in fact
complementary. The $\tau$ result given here probes the coupling to
a transverse W (helicity $\pm$1) while the $\pi$ decays measure the
coupling to a longitudinal W (helicity 0). It is indeed conceivable
that either approach could be sensitive to different nonstandard
corrections to universality.

Since $B_e$ and $B_\mu$ are consistent with $\mu$-e universality their
values can be combined, taking common errors into account, into 
a consistent leptonic branching ratio for a massless lepton (essentially
the case for the electron, noting that $f(\frac {m^2_e}{m^2_\tau}) = 1$
for all practical purposes)
\beq
B^{(m_l=0)}_l = 17.822 \pm 0.044 \pm 0.022 (\%)~,
\eeq
where the first error is statistical and the second from systematic effects.

\subsubsection{Tests of $\tau$-$\mu$ and  $\tau$-e universality}
\label{univ}

Comparing the rates for 
$\Gamma(\tau \rightarrow \nu_\tau e \overline{\nu}_e (\gamma))$,
$\Gamma(\tau \rightarrow \nu_\tau \mu \overline{\nu}_\mu (\gamma))$ 
and $\Gamma(\mu \rightarrow \nu_\mu e \overline{\nu}_e (\gamma))$
provides direct tests of the universality of $\tau$-$\mu$-e couplings.
Taking the relevant ratios with calculated radiative corrections, 
one obtains
\beqn
 \left(\frac {g_\tau}{g_\mu}\right)^2 &=& \frac {\tau_\mu}{\tau_\tau} 
 \left(\frac{m_\mu}{m_\tau}\right)^5 B_e
 \frac {f(\frac {m^2_e}{m^2_\mu})}{f(\frac {m^2_e}{m^2_\tau})} 
 \Delta_W \Delta_\gamma \\
 \left(\frac {g_\tau}{g_e}\right)^2 &=& \frac {\tau_\mu}{\tau_\tau} 
 \left(\frac{m_\mu}{m_\tau}\right)^5 B_\mu 
 \frac {f(\frac {m^2_e}{m^2_\mu})}{f(\frac {m^2_\mu}{m^2_\tau})} 
 \Delta_W \Delta_\gamma
\eeqn 
where $f(\frac {m^2_e}{m^2_\mu}) = 0.9998$,
$\Delta_W = \frac {\delta^\mu_W}{\delta^\tau_W} = 1 - 2.9\cdot 10^{-4}$,
$\Delta_\gamma = \frac {\delta^\mu_\gamma}{\delta^\tau_\gamma} 
= 1 + 8.5\cdot 10^{-5}$,
and $\tau_l$ is the lepton $l$ lifetime.

From the present measurements of $B_e$, $B_\mu$, the $\tau$ 
mass~\cite{pdg2000}, $m_\tau = (1777.03^{+0.30} _{-0.26})$~MeV (dominated
by the BES result~\cite{besmtau}), the $\tau$ lifetime~\cite{pdg2000},
$\tau_\tau = (290.6 \pm 1.1)$ fs and the other 
quantities from Ref.~\cite{pdg2000}, universality can be tested:
\beqn
 \frac {g_\tau}{g_\mu} &=& 
 1.0009 \pm 0.0023 \pm 0.0019 \pm 0.0004 \\
 \frac {g_\tau}{g_e} &=& 
 1.0001 \pm 0.0022 \pm 0.0019 \pm 0.0004~,
\eeqn
where the errors are from the corresponding leptonic branching ratio
and the $\tau$ lifetime and mass, respectively.
\subsubsection{$\tau$-$\mu$ universality from the pionic branching ratio}

The measurement of $B_\pi$ also permits an independent test of 
$\tau$-$\mu$ universality through the relation
\beqn
 \left(\frac{g_\tau}{g_\mu}\right)^{\!\!2} &=& 
     \frac{B_\pi}{B_{\pi \rightarrow \mu \overline{\nu}_\mu}}
     \frac{\tau_\pi}{\tau_\tau} \frac{2 m_\pi m_\mu^2}{m_\tau^3} \nonumber\\
	&& \hspace{0.0cm}
     \times\;\left(\frac{1-m_\mu^2/m_\pi^2}
                {1-m_\pi^2/m_\tau^2}\right)^{\!\!2}
      \delta_{\tau / \pi}~,
\eeqn
where the radiative correction~\cite{decker-fink} amounts to
$\delta_{\tau / \pi} = 1.0016 \pm 0.0014$
Using the world-averaged values for the $\tau$ and $\pi$ ($\tau_\tau$ 
and $\tau_\pi$) lifetimes, and the branching ratio for the decay
$\pi \rightarrow \mu \nu$~\cite{pdg2000}, the present result
for $B_\pi$, one obtains
\beq
\label{eq_taumu_pi}
  \frac {g_\tau} {g_\mu} = 
0.9962 \pm 0.0048 \pm 0.0019 \pm 0.0002~,
\eeq
comparing the measured value ($B_\pi = 10.823 \pm 0.104$)\% 
to the expected one assuming universality ($10.910 \pm 0.042$)\%.
The quoted errors in Eq.~\ref{eq_taumu_pi} are from the pion mode branching
ratio and the $\tau$ lifetime and mass, respectively.

The two determinations of $\frac {g_\tau}{g_\mu}$ obtained from $B_e$
and $B_\pi$ are consistent with each other and can be combined to yield
\beq
  \frac {g_\tau} {g_\mu} = 
1.0000 \pm 0.0021 \pm 0.0019 \pm 0.0004~,
\eeq
where the errors are from the electron and pion branching ratio
and the $\tau$ lifetime and mass, respectively.
Universality of the $\tau$ and $\mu$ charged-current couplings holds
at the 0.29\% level with about equal contributions from the present 
determination of $B_e$ and $B_\pi$, and the world-averaged value 
for the $\tau$ lifetime.

\begin{figure}[t]
   \centerline{\psfig{file=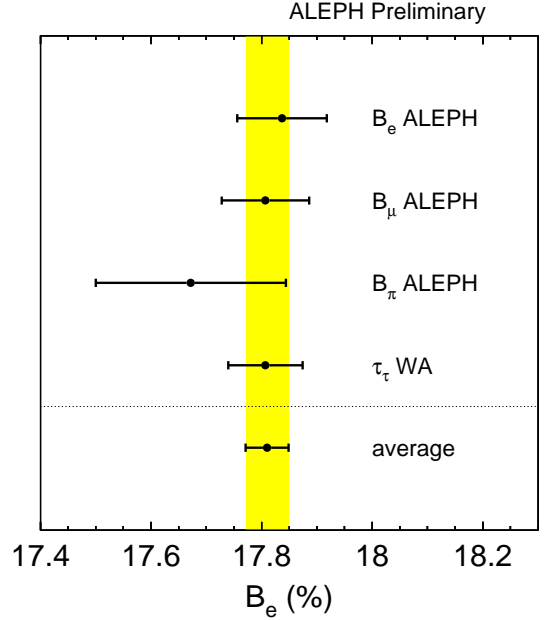,width=70mm}}
   \caption{The measured value for $B_e$ compared to predictions from
  other measurements assuming leptonic universality. The vertical band
  gives the average of all determinations.}
   \label{allbe}
\end{figure}

The consistency of the present branching ratio measurements with leptonic
universality is displayed in Fig.~\ref{allbe} where the result for $B_e$
is compared to computed values of $B_e$ using as input $B_\mu$ (assuming
$e - \mu$ universality), $\tau_\tau$ and $\tau_\mu$ ($\mu - \tau$ 
universality), and $B_\pi$ and $\tau_\pi$ ($\mu - \tau$ universality). 
All values are consistent and yield the average
\beq
 B_e^{\rm universality} = (17.810 \pm 0.039)~\%~.
\eeq

\subsubsection{$a_1$ decays to $3 \pi$ and $\pi 2 \pi^0$}
\label{a1}

With the level of precision reached it is interesting to compare the rates
in the $3 \pi$ and $\pi 2 \pi^0$ channels which are completely dominated
by the $a_1$ resonance. The dominant $\rho \pi$ intermediate state leads
to equal rates, but a small isospin-breaking effect is expected from
different charged and neutral $\pi$ masses, slightly favouring the  
$\pi 2 \pi^0$ channel.

A recent CLEO partial-wave analysis of the $\pi 2 \pi^0$ final 
state~\cite{cleoa1} has shown that the situation is in fact much more 
complicated with many intermediate states, in particular involving 
isoscalars, amounting to about 20\% of the total rate and producing 
strong interference effects. A good description of the $a_1$ decays 
was achieved in the CLEO study, which can be applied to the $3 \pi$ final 
state, predicting~\cite{cleoa1} a ratio of the rates $3 \pi$/$\pi 2 \pi^0$
equal to 0.985. This value, which includes known isospin-breaking from
the pion masses, turns out to be in good agreement with the measured value
from this analysis which shows the expected trend
\beq
 \frac{B_{3 \pi}}{B_{\pi 2 \pi^0}} = 0.979 \pm 0.018~.
\eeq

\subsubsection{The $\pi\piz$ branching ratio in the context 
of $a_\mu^{had}$}

The $\pi\piz$ final state is dominated by the $\rho$ resonance as 
demonstrated in Fig.~\ref{spec_mass}. Its mass
distribution ---or better the corresponding spectral function--- 
is a basic ingredient of vacuum polarization calculations,
such as that used for computing the hadronic contribution to the
anomalous magnetic moment of the muon $a_\mu^{had}$. In this case the
$\rho$ contribution is dominant (71\%) and therefore controls the final 
precision of the result. The current evaluation~\cite{dh} used by the
BNL experiment~\cite{bnl02} is based on $\tau$ and $e^+ e^-$ data, 
with more precision on the $\tau$ side.

The normalization of the spectral function is provided by the
branching fraction $B_{\pi\piz}$ taken from the world average
and completely dominated by the published ALEPH result~\cite{aleph13_h}. 
The new result given here is larger by 0.68\%, thus one can expect
a slightly larger contribution to $a_\mu^{had}$, but it remains within
the quoted uncertainty~\cite{dh} of $6.2\cdot 10^{-10}$, corresponding
to a relative error of 1.2\% for the $\rho$ contribution.

\begin{figure}[t]
   \centerline{\psfig{file=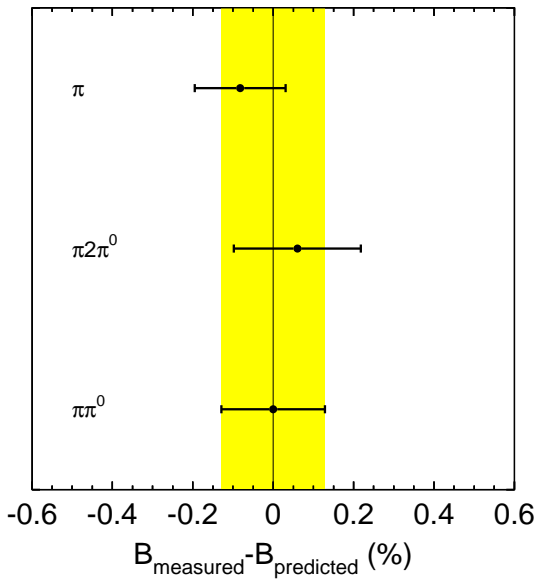,width=70mm}}
   \caption{The differences between the measured values for $B_{\pi}$ and
  $B_{\pi 2\piz}$ and their respective predictions from leptonic universality
  and isospin symmetry with $\pi$-mass breaking, compared to the precision
  on $B_{\pi \piz}$.}
   \label{brho_check}
\end{figure}

A new evaluation~\cite{dehz,andreas02} is available, using the spectral 
functions from the present preliminary ALEPH analysis, the published CLEO 
results~\cite{cleo_2pi} and new results from $e^+e^-$ annihilation 
from CMD-2~\cite{cmd2}. Since a disagreement is observed between the
$\tau$ and $e^+e^-$ spectral functions, it is important to check all
ingredients, in particular the determination of the branching ratio
$B_{\pi\piz}$. As most of the systematic uncertainty in $B_{\pi\piz}$ 
comes from $\gamma / \piz$ reconstruction, it is interesting to cross
check the results in the 'adjacent' hadronic modes, {\it i.e.} the
$\pi$ and $\pi 2\piz$ channels. This is possible if universality in the 
weak charged current is assumed, leading to an absolute prediction of
$B_{\pi}$ using as input the $\tau$ lifetime (see Section~\ref{univ}),
and by computing $B_{\pi 2\piz}$ from the measurement of $B_{3\pi}$ which
is essentially uncorrelated with $B_{\pi\piz}$ (see Section~\ref{a1}).
The comparison, shown in Fig.~\ref{brho_check}, does not point to any
systematic bias in the determination of $B_{\pi\piz}$ within the quoted 
uncertainty.

\subsubsection{Separation of vector and axial-vector contributions}
\label{vasep}

From the complete analysis of the $\tau$ branching ratios presented in
this paper, it is possible to determine the nonstrange vector (V) 
and axial-vector (A) contributions to the total $\tau$ hadronic width, 
conveniently expressed in terms of their ratios to the leptonic width, 
called $R_{\tau,V}$ and $R_{\tau,A}$, respectively. The determination of 
the strange counterpart $R_{\tau,S}$ is already published~\cite{alephksum}.

The ratio \Rt\ for the total hadronic width
is calculated from the dif\/ference of the ratio of 
the total hadronic width and electronic \br,
\beqn
  R_\tau&=&\frac{1-B_e-B_\mu}{B_e}
           =\frac{1}{B_e}-1.97257
          \nonumber \\
        &=& 3.642 \pm 0.012~. 
\eeqn
taking for $B(\tau^-\rightarrow e^-\,\bar{\nu}_e\nu_\tau)$ the value obtained
in Section~\ref{uni} assuming universality in the leptonic weak current.
Using the ALEPH measurement of the strange width ratio~\cite{alephksum},
very slightly modified to take into account the channel $K^{*-} \eta$
measured by CLEO~\cite{cleoksteta} 
\beq
   R_{\tau,S} = 0.1625\pm0.0066~,
\eeq
the following result is obtained for the nonstrange component
\beq
\label{eq_rtauvpa}
    R_{\tau,V+A}    = 3.480 \,\pm\, 0.014~.
\eeq

Separation of V and A components in hadronic final states with only pions
is straightforward. Isospin invariance relates the spin-parity of such 
systems to their number of pions: G-parity =1 (even number) corresponds to
vector states, while G=-1 (odd number) tags axial-vector states. This
property places a strong requirement on the efficiency of $\piz$
reconstruction, a constraint that was strongly integrated in this analysis.

Modes with a $K \overline{K}$ pair are not in general eigenstates 
of G-parity and contribute to both V and A channels. The respective
components have been determined in the ALEPH analysis~\cite{alephksum}.
While the decay to $K^- K^0$ is pure vector, the 
$K \overline{K} \pi$ mode has been shown to be almost completely
axial vector, with a fraction $0.94^{+0.06}_{-0.08}$. This information
was not available at the time of the previous analysis of the nonstrange
modes~\cite{aleph13_h} where a conservative value of $0.5 \pm 0.5$ was
used. For the decays into $K \overline{K}\pi\pi$ no information is
available in this respect and the same conservative fraction is assumed.

The total nonstrange vector and axial-vector contributions 
obtained in this analysis are:
\beqn
\label{eq_rtauv}
  R_{\tau,V}&=& 1.778 \pm 0.010 \pm 0.002~,\\
\label{eq_rtaua}
  R_{\tau,A}&=& 1.701 \pm 0.011 \pm 0.002~,
\eeqn
where the second errors reflect the uncertainties in the $V/A$ separation
in the channels with $K \overline{K}$ pairs.
Taking care of the correlations between the respective uncertainties, one
obtains the difference between the vector and axial-vector components,
which is physically related to the amount of nonperturbative QCD 
contributions in the nonstrange hadronic $\tau$ decay width:
\beq
\label{eq_rtauvma}
    R_{\tau,V-A}    = 0.077 \pm 0.018 \pm 0.005~,
\eeq
where again the second error has the same meaning 
as in Eqs.~(\ref{eq_rtauv}) and (\ref{eq_rtaua}). The ratio
\beq
\label{eq_rvmavpa}
    \frac {R_{\tau,V-A}} {R_{\tau,V+A}}     = 0.022 \pm 0.005
\eeq
is a measure of the relative importance of nonperturbative QCD contributions.

\section{Conclusions}

A final analysis of $\tau$ decay branching fractions using all LEP-I
data with the ALEPH detector is presented. As in the publication based on
the 1991-1993 data it uses a global analysis of all modes, classified 
according to charged particle identification, and charged particle and 
$\pi^0$ multiplicity up to 4 $\piz$s in the final state. Major improvements 
are introduced with respect to the published analysis and a better 
understanding is achieved, in particular in the separation between 
genuine and fake photons. In this process shortcomings and small biases
of the previous method were discovered are corrected, leading to more 
robust results. As modes with kaons ($K^\pm$, $K^0_S$, and $K^0_L$) 
have already been studied and published with the full statistics, 
the nonstrange modes without kaons are emphasized. Taken together 
these results provide a complete description of $\tau$ decay modes 
up to 6 hadrons in the final state.

\begin{table*} 
\caption{A summary list of ALEPH branching ratios (\%). The labels $V$,
 $A$ and $S$ refer to the nonstrange vector and axialvector, and strange
components, respectively.}
\begin{center}
\setlength{\tabcolsep}{1.5pc}
\begin{tabular}{lrrc}
\hline\hline
 mode & B $\pm\sigma_{\hbox{tot}}$
   [\%] & {(PRELIMINARY!)}\\\hline
   \phyi &    17.837 $\pm$  0.080 &&\\
   \phyii &    17.319 $\pm$  0.077 &&\\
\hline
   \phyiii &    10.828 $\pm$  0.105 &A&\\
   \phyiv &    25.471 $\pm$  0.129  &V&\\
   \phyv &     9.239 $\pm$  0.124  &A&\\
   \phyvi &      0.977 $\pm$  0.090  &V&\\
   \phyxiii &      0.112 $\pm$  0.051  &A&\\
   \phyvii &     9.041 $\pm$  0.097  &A&\\
   \phyviii &     4.590 $\pm$  0.086  &V&\\
   \phyix &      0.392 $\pm$  0.046  &A&\\
   \phyx &      0.013 $\pm$  0.010  &V &estim\\
   \phyxi &      0.072 $\pm$  0.015  &A&\\
   \phyxii &      0.014 $\pm$  0.009  &V&\\
   $\pi^- \pi^0 \eta$ & 0.180 $\pm$ 0.045  &V& \\
   $(3\pi)^-  \eta$ & 0.039 $\pm$ 0.007  &A& CLEO\\
   $a_1^- (\rightarrow \pi^- \gamma)$ & 0.040 $\pm$ 0.020  &A& estim \\
   $\pi^- \omega (\rightarrow \pi^0 \gamma, \pi^+ \pi^-)$ & 0.253 $\pm$ 0.018     &V& \\
   $\pi^- \pi^0 \omega (\rightarrow \pi^0 \gamma, \pi^+ \pi^-)$ & 0.048 $\pm$
   0.009  &A& + CLEO\\
   $(3\pi)^- \omega (\rightarrow \pi^0 \gamma, \pi^+ \pi^-)$ & 0.003 $\pm$
   0.003  &V& CLEO\\
\hline
   $K^- K^0 $ &  0.163 $\pm$ 0.027 &V&\\
   $K^- \pi^0 K^0 $    &  0.145 $\pm$ 0.027  & $(94^{+6}_{-8})\%$A&\\
   $\pi^- K^0 \overline{K^0}$
                &  0.153 $\pm$ 0.035  &$(94^{+6}_{-8})\%$A&\\
   $K^- K^+ \pi^- $    &  0.163 $\pm$ 0.027  &$(94^{+6}_{-8})\%$A&\\
   $(K\overline{K}\pi\pi)^-$ &  0.05  $\pm$ 0.02   &$(50 \pm 50)\%$ A&\\
\hline
   $K^- $ &  0.696 $\pm$ 0.029 &S&\\
   $K^- \pi^0 $    &  0.444 $\pm$ 0.035  &S&\\
   $\overline{K^0} \pi^-$ &  0.917 $\pm$ 0.052  &S&\\
   $K^- 2\pi^0 $    &  0.056 $\pm$ 0.025  &S&\\ 
   $K^- \pi^+ \pi^- $    &  0.214 $\pm$ 0.047  &S&\\
   $\overline{K^0} \pi^- \piz$ &  0.327 $\pm$ 0.051  &S&\\
   $(K 3\pi)^- $ &  0.076 $\pm$ 0.044 &S&\\
   $K^- \eta$  &  0.029 $\pm$ 0.014 &S&\\
\hline\hline
\end{tabular}
\label{completeBR}
\end{center}
\end{table*}

The measured branching ratio values are internally consistent and agree with
known constraints from other measurements in the framework of the
Standard Model (or even looser assumptions). The precision reached and the
completeness of the results are for the moment unique. More specifically,
the results on the leptonic and pionic fractions lead to powerful tests 
of universality in the charged leptonic weak current, showing that 
the $e-\mu-\tau$ couplings are equal within 2-3 per mille. The branching
ratio of $\tau \rightarrow \nu_\tau \pi \piz$ which is of particular interest
to the accurate determination of vacuum polarization effects is determined
with a precision of 0.5\% to be $(25.47 \pm 0.13)~\%$. Also the ratio of $a_1$
branching fractions into $\pi 2\piz$ and $3\pi$ final states is measured
to be $0.979 \pm 0.018$, in agreement with expectation from partial wave
analyses of these decays. Separating nonstrange hadronic channels into
vector (V) and axial-vector (A) components and normalizing to the electronic
width yields the ratios $R_{\tau,V} = 1.778 \pm 0.010$, 
$R_{\tau,A} = 1.701 \pm 0.011$, $R_{\tau,V+A} = 3.480 \pm 0.014$ and 
$R_{\tau,V-A} = 0.077 \pm 0.019$.

The results presented here are combined with previously published ALEPH 
results on final states with kaons in Table~\ref{completeBR}.

\section*{Acknowledgements}

We would like to thank the organizers of the Tau02 Workshop for 
their hospitality and their efficient running of the meeting.

\end{document}